\documentclass[aps,prl,preprint,superscriptaddress]{revtex4}
\usepackage{graphicx}
\usepackage{color}
\usepackage{amsfonts}
\usepackage{ulem}

\definecolor{brown}{RGB}{200,100,0}

\newcommand{\EF}{$E_\mathrm{F}$}

\def\kbar{$\bar{\mathrm{K}}$}
\def\kbarp{\kbar$^{\prime}$}
\def\gbar{$\bar{\mathrm{\Gamma}}$}
\def\kgk{\kbar-\gbar-\kbarp}
\def\WS2{WS$_2$}
\def\MoS2{MoS$_2$}{

\def\EF{$E\mathrm{_F}$}

\def\xplus{$X^+$}
\def\xminus{$X^-$}

\def\figone{\begin{figure*}
\begin{center}
\includegraphics[width=0.95\textwidth]{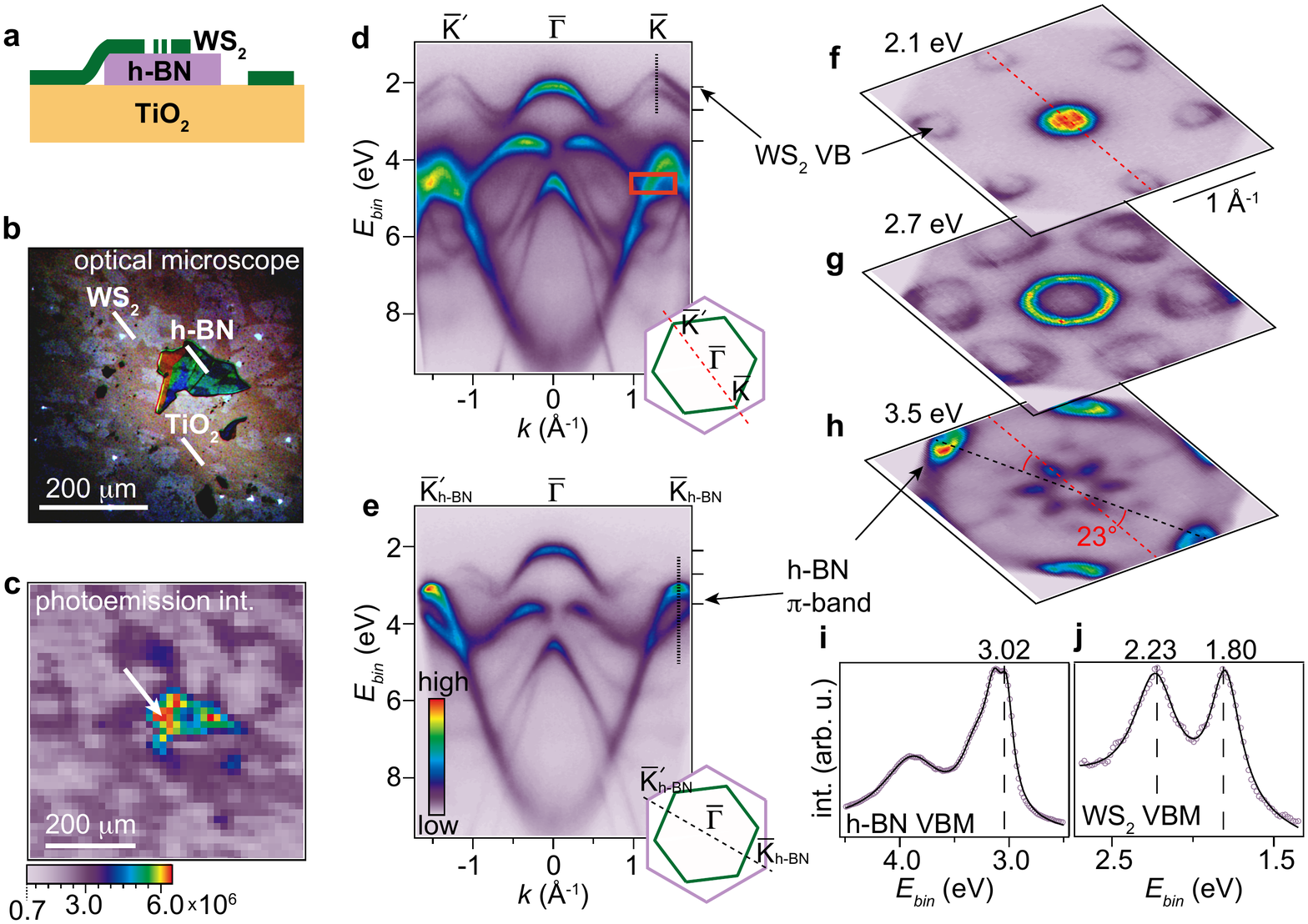}
\caption{\textbf{Spatially-resolved electronic structure mapping of a WS$_2$/h-BN heterostructure supported on TiO$_2$.} \textbf{a,} Side-view sketch of WS$_2$/h-BN on TiO$_2$, illustrating SL WS$_2$ regions contacted directly to h-BN and to TiO$_2$. \textbf{b,} Optical microscope image of the sample. The contrast has been strongly enhanced to better visualize the SL WS$_2$. Brown patches correspond to bare TiO$_2$, light purple to WS$_2$/TiO$_2$ and the darker green/red structure is the h-BN flake. \textbf{c,} Spatial map of photoemission intensity (integrated over the red box in (d)) for the same region seen in \textbf{b}. See Supplementary Section 1 for details on the spatial intensity variations. \textbf{d,} Measured dispersion along the \kgk\ direction of the SL WS$_2$ BZ (see green BZ and dashed red line in the insert) collected at the spatial coordinates marked by a white arrow in \textbf{c}. The rectangular red box marks a region with crossing WS$_2$ and h-BN bands where the photoemission intensity is integrated to produce the spatial map in \textbf{c}. \textbf{e,} ARPES dispersion in the high symmetry direction of h-BN (see purple BZ and dashed black line in the insert). \textbf{f-h,} Constant energy cuts obtained at the given binding energies (see also ticks on the right of panels \textbf{d}-\textbf{e}). Arrows mark distinct energy contours relating to SL WS$_2$ and to h-BN. The red and black dashed lines (insert in \textbf{d-e}) indicate a twist angle of (23 $\pm$ 1)$^{\circ}$ in between the SL WS$_2$ and h-BN. \textbf{i-j,} EDCs obtained along the dotted lines in \textbf{d-e} around the h-BN VBM (\textbf{i}) and SL WS$_2$ VBM  (\textbf{j}). The positions of the band edges are given in units of eV and the error bar is 30~meV.}
\label{fig:1}
\end{center}
\end{figure*}}

\def\figtwo{\begin{figure*}
\begin{center}
\includegraphics[width=0.75\textwidth]{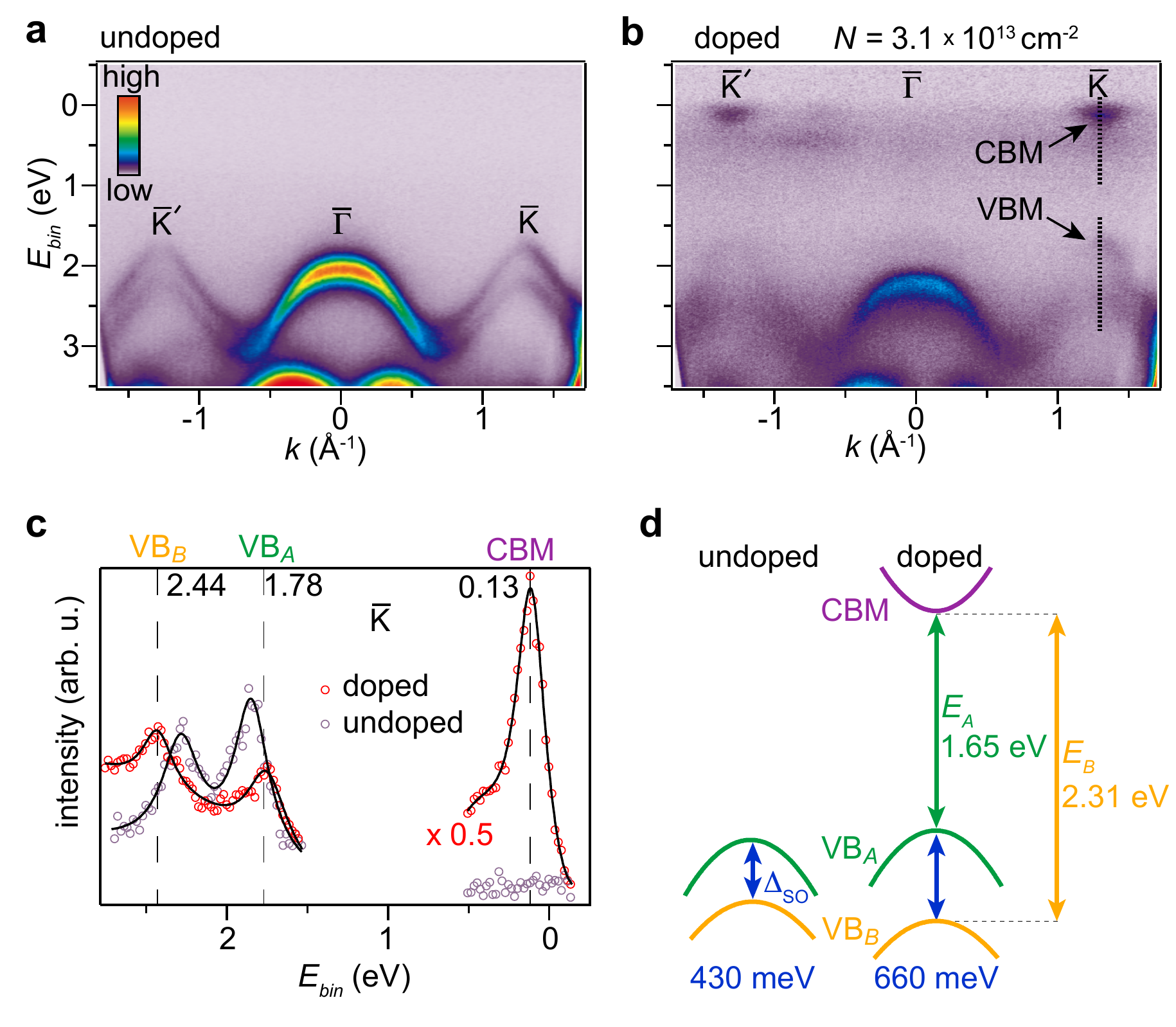}
\caption{\textbf{Electronic structure of strongly electron-doped WS$_2$/h-BN.} \textbf{a,} Dispersion of SL WS$_2$/h-BN along the \kgk\ direction of the SL WS$_2$ BZ, collected from the spot marked by the white arrow in Fig.~\ref{fig:1}(c).  \textbf{b,} Corresponding data at the highest achieved electron-doping $N$. \textbf{c,} EDCs (markers) around the VBM and CBM at \kbar\ (see dotted lines and arrows in \textbf{b}) for undoped and electron-doped SL WS$_2$ on h-BN. Peak positions extracted from Lorentzian line fits (curves) are shown as vertical dashed lines and values for the doped case are given in units of eV. Note that the CBM intensity has been scaled by a factor of 0.5 in order to make the comparison with the other peaks more clear. \textbf{d,} Schematics of the dispersion change of the VB spin-orbit split bands VB$_A$ and VB$_B$ due to doping and the measured energy gaps between the CBM and VB$_A$ (denoted as $E_A$) and VB$_B$ (denoted as $E_B$) in the doped case. The energy splitting due to spin-orbit coupling $\Delta_{SO}$ and the values of $E_A$ and $E_B$ are provided with error bars of 30~meV.}
\label{fig:2}
\end{center}
\end{figure*}}

\def\figthree{\begin{figure*}
\begin{center}
\includegraphics[width=0.95\textwidth]{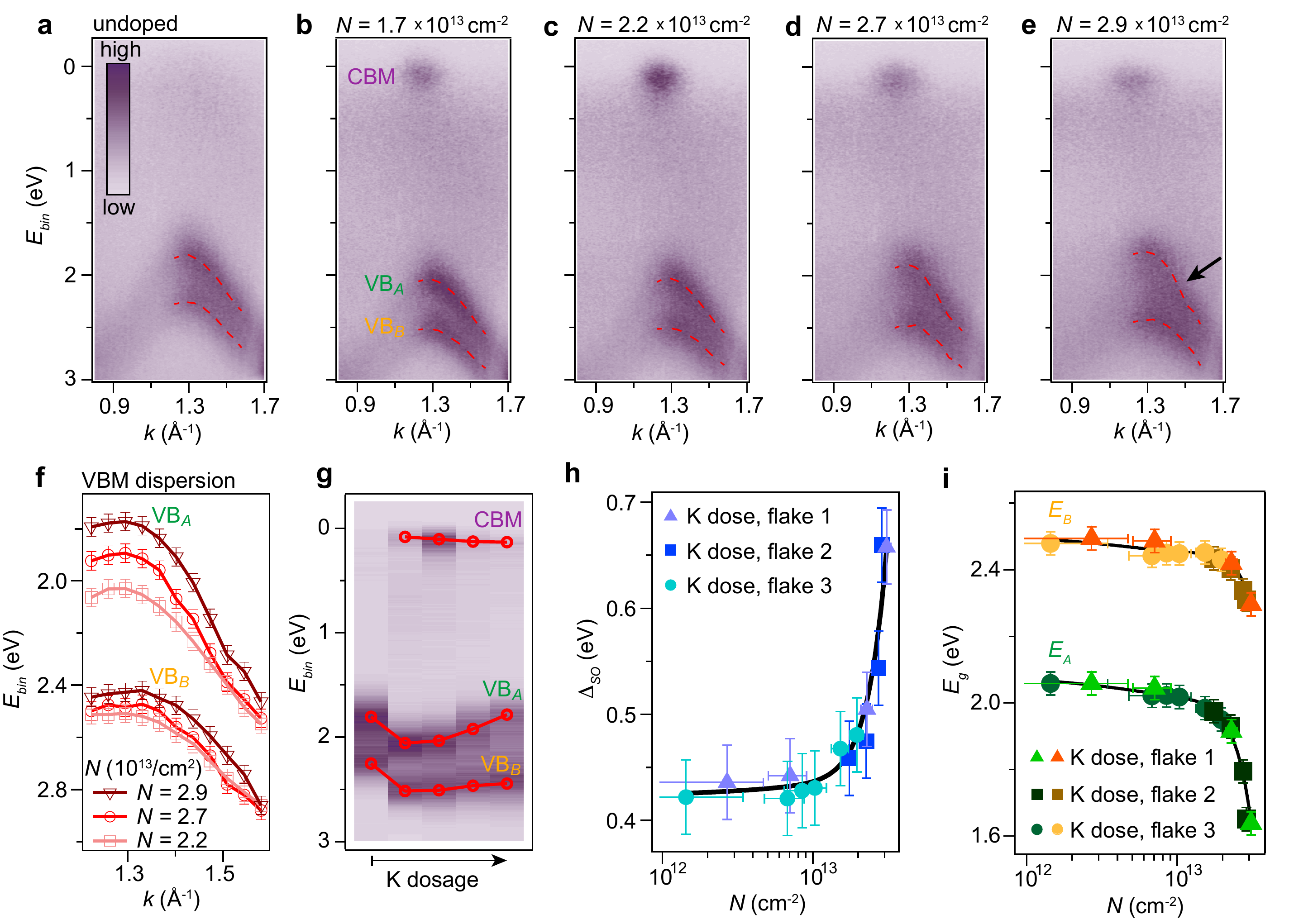}
\caption{\textbf{Evolution of SL WS$_2$ VBM and CBM dispersion with charge carrier density.} \textbf{a-e,} microARPES measurements around the \kbar-point of SL WS$_2$/h-BN for clean (\textbf{a}) and increasingly potassium-dosed cases \textbf{(b)-(e)}. The red curves are the fitted VB$_A$ and VB$_B$ dispersions determined via double Lorentzian fits of EDCs (see Supplementary Figure S6). The arrow in \textbf{e} points to a possible kink in VB$_A$. \textbf{f,} VB$_A$ and VB$_B$ dispersions extracted from the EDC analysis in the strongly electron doped cases. \textbf{g,} ARPES intensity at the $\bar{K}$-point and peak positions (red markers) at each potassium dosing step (see corresponding EDC analysis in Supplementary Figure S7). \textbf{h-i,} Spin-orbit splitting (\textbf{h}) and band gap values (\textbf{i}) determined from the VB$_A$, VB$_B$ and CBM positions, combining data from K doping experiments on the different WS$_2$/h-BN flakes investigated here, in Fig. 2 and in Supplementary Figure S8. The lines in \textbf{h-i} are provided as guides to the eye.}
\label{fig:3}
\end{center}
\end{figure*}}

\def\figfour{\begin{figure*}
\begin{center}
\includegraphics[width=0.7\textwidth]{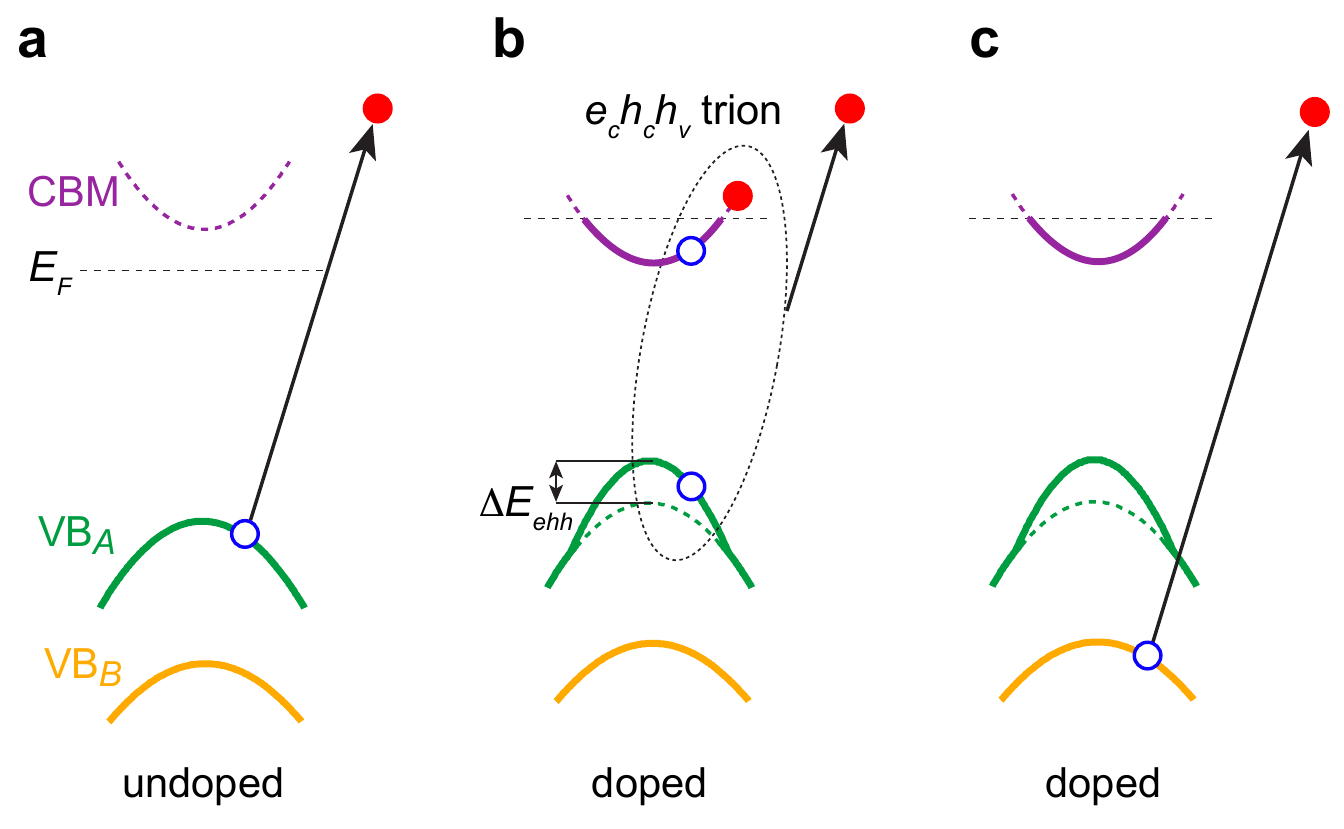}
\caption{\textbf{Quasiparticle dynamics in doped SL WS$_2$.} \textbf{a,} Band diagram illustrating the generated photohole (hollow blue circle) in VB$_A$ and the photoemitted free electron (filled red circle) in the undoped situation. \textbf{b,} Illustration of an $e_ch_ch_v$ trionic quasiparticle generated in doped SL WS$_2$. The photohole binding energy is lowered and the VB$_A$ dispersion renormalizes with respect to the bare band (green dashed curve) by the trion binding energy $\Delta E_{ehh}$. \textbf{c,} A photohole generated in VB$_B$ in the doped situation undergoes a process that is analogous to the undoped situation in \textbf{a} due to the absence of strong trion interactions with photoholes in this band.}
\label{fig:4}
\end{center}
\end{figure*}}

\begin{document}
\title{Giant spin-splitting and gap renormalization driven by trions in single-layer WS$_2$/h-BN heterostructures}
\author{ Jyoti Katoch$^{\ast}$}
\affiliation{ Department of Physics, The Ohio State University, Columbus, Ohio 43210, USA}
\author{ S{\o}ren Ulstrup$^{\ast}$}
\affiliation{ Advanced Light Source, E. O. Lawrence Berkeley National Laboratory, Berkeley, California 94720, USA}
\affiliation{Department of Physics and Astronomy, Interdisciplinary Nanoscience Center (iNANO), Aarhus University, 8000 Aarhus C, Denmark}
\author{ Roland J. Koch}
\affiliation{ Advanced Light Source, E. O. Lawrence Berkeley National Laboratory, Berkeley, California 94720, USA}
\author{ Simon Moser}
\affiliation{ Advanced Light Source, E. O. Lawrence Berkeley National Laboratory, Berkeley, California 94720, USA}
\author{ Kathleen M. McCreary}
\affiliation{Naval Research laboratory, Washington, D.C. 20375, USA}
\author{ Simranjeet Singh}
\affiliation{ Department of Physics, The Ohio State University, Columbus, Ohio 43210, USA}
\author{ Jinsong Xu}
\affiliation{ Department of Physics, The Ohio State University, Columbus, Ohio 43210, USA}
\author{ Berend T. Jonker}
\affiliation{Naval Research laboratory, Washington, D.C. 20375, USA}
\author{ Roland K. Kawakami }
\affiliation{ Department of Physics, The Ohio State University, Columbus, Ohio 43210, USA}
\author{ Aaron Bostwick}
\affiliation{ Advanced Light Source, E. O. Lawrence Berkeley National Laboratory, Berkeley, California 94720, USA}
\author{ Eli Rotenberg}
\affiliation{ Advanced Light Source, E. O. Lawrence Berkeley National Laboratory, Berkeley, California 94720, USA}
\author{ Chris Jozwiak$^\dagger$}
\affiliation{ Advanced Light Source, E. O. Lawrence Berkeley National Laboratory, Berkeley, California 94720, USA} 
\affiliation{ $^{\ast}$ These authors contributed equally to the work. \\ $^\dagger$ Email: CMJozwiak@lbl.gov}

\maketitle 
\textbf{In two-dimensional (2D) semiconducting transition metal dichalcogenides (TMDs), new electronic phenomena such as tunable band gaps \cite{Chernikov:2015b,Liang:2015,Chernikov:2015} and strongly bound excitons and trions emerge from strong many-body effects \cite{ramasl2012,Qiu:2013,Mak:2013}, beyond spin-orbit coupling- and lattice symmetry-induced spin and valley degrees of freedom \cite{Xiao:2012}. Combining single-layer (SL) TMDs with other 2D materials in van der Waals heterostructures offers an intriguing means of controlling the electronic properties through these many-body effects via engineered interlayer interactions \cite{Geim:2013,Ugeda:2014,Larentis:2014}. Here, we employ micro-focused angle-resolved photoemission spectroscopy (microARPES) and \textit{in-situ} surface doping to manipulate the electronic structure of SL WS$_2$ on hexagonal boron nitride (WS$_2$/h-BN). Upon electron doping, we observe an unexpected giant renormalization of the SL WS$_2$ valence band (VB) spin-orbit splitting from 430~meV to 660~meV, together with a band gap reduction of at least 325~meV, attributed to the formation of trionic quasiparticles. These findings suggest that the electronic, spintronic and excitonic properties are widely tunable in 2D TMD/h-BN heterostructures, as these are intimately linked to the quasiparticle dynamics of the materials \cite{Xu:2014,Kormnyos:2015,Mak:2016}.}

Coulomb interactions in 2D materials are several times stronger than in their 3D counterparts. In 2D TMDs, this is most directly evidenced by the presence of excitons with an order of magnitude higher binding energies than in the bulk \cite{ramasl2012}. While the excitons in these 2D materials have been widely studied by optical techniques \cite{Mak:2016}, the impact of strong electron-electron interactions on the quasiparticle band structure remains unclear. Theory predicts many-body effects to influence the spin-orbit splitting around the valence band maximum (VBM) and conduction band minimum (CBM) \cite{Ferreiros:2014}. While these should be observable by ARPES, a direct probe of many-body effects \cite{Bostwick:2007}, measurements so far have mainly focused on the layer-dependence of the single-particle spectrum and the direct band gap transition in 2D TMD systems, including epitaxial SL MoSe$_2$ \cite{Zhang:2014a} and SL WSe$_2$ \cite{Zhang:2016} grown on doped multilayer graphene, and SL MoS$_2$ grown on a metal surface \cite{Miwa:2015}. On such conductive substrates the interfacial interactions and screening are known to strongly influence the electronic properties of the SL TMD \cite{Ugeda:2014}. 
\figone

Flakes of SL TMDs have been transferred on oxide substrates such as SiO$_2$ where the substrate screening and interfacial effects are potentially reduced. However, resulting ARPES spectra have been too broad for detailed analysis \cite{Jin2013}, likely due to large surface roughness and charge impurity scattering \cite{Dean:2010}. With respect to SiO$_2$ and similar substrates, h-BN has favorable qualities like atomic flatness, modest screening and a homogeneous charge distribution. This should enable direct investigation of the adjacent TMD's intrinsic electronic structure and many-body effects. h-BN is often used as a substrate for graphene heterostructures \cite{Dean:2010,Geim:2013} with high device performance \cite{Wang:2013} and new exotic electronic states such as quantized Dirac cones \cite{Wang:2016b}. Unfortunately, the lateral size of mechanically assembled heterostructures is usually on the order of $\sim 10$~$\mu$m, much smaller than the beam spot of typical ARPES setups ($\gtrsim$100~$\mu$m). Furthermore, sample charging on insulating bulk h-BN substrates would complicate ARPES experiments.

We overcome these challenges as follows. We realize a high quality 2D semiconductor-insulator interface by mechanical transfer of a relatively large ($\sim$100~$\mu$m) SL WS$_2$ crystal over a thin flake of h-BN that was transferred onto a degenerately doped TiO$_2$ substrate, as depicted in Fig.~\ref{fig:1}(a). Sample charging is avoided by electrically contacting the continuous SL WS$_2$ flake to both the h-BN and the conductive TiO$_2$. Fig. \ref{fig:1}(b) is an optical microscope image of the sample, including a $\approx 100$~$\mu$m wide h-BN flake, surrounded by several transferred flakes of SL WS$_2$ on the TiO$_2$ substrate (WS$_2$/TiO$_2$), one of which partially overlaps the h-BN.

By using a state-of-the-art spatially-resolved microARPES experiment with a 10~$\mu$m focused synchrotron beam spot, we are able to collect distinct high quality band structure information from the multiple micron-scale interfaces. A spatial map of the photoemission intensity around the WS$_2$/h-BN heterostructure is shown in Fig. \ref{fig:1}(c), which was produced by integrating the intensity over the boxed region of the corresponding $k$-space band structure shown in Fig. \ref{fig:1}(d), measured at each spatial point. 
The crossing SL WS$_2$ and h-BN bands in this region ensure strong contrast between WS$_2$/h-BN (white arrow), regions of WS$_2$/TiO$_2$ (light purple), and regions of bare TiO$_2$ (dark purple) in the spatial map.
The photoemission map (panel (c)) corresponds directly to the optical micrograph (panel (b)) with contrasts that reflect the intensity of the WS$_2$ and h-BN features in the red box in panel (d). The band structures from bare TiO$_2$, WS$_2$/TiO$_2$ and several spots within the  WS$_2$/h-BN heterostructure are presented in Supplementary Figure S1. The slight intensity variations within the WS$_2$/h-BN heterostructure arise from areas with pinholes introduced in the SL WS$_2$ during transfer, as sketched in Fig. \ref{fig:1}(a) \cite{Ulstrup:2016}. The sensitivity towards such features, which are not resolved by the optical micrograph, demonstrates the capability of identifying optimum sample areas directly in the ARPES experiment, which is critical for such complex, heterogenous samples.
\figtwo

The VB electronic structure through the entire first Brillouin zone (BZ) of the heterostructure, including the SL WS$_2$ bands and the $\pi$-band dispersion of the underlying h-BN, is shown in Figs. \ref{fig:1}(d)-(e). The data is collected from a single spatial point where the WS$_2$ features are most intense (white arrow in Fig. \ref{fig:1}(c)). Custom electrostatic deflectors mounted in the photoelectron analyzer enable full scans of $k$-space at exactly this position without any drift from sample motion. The BZ orientations and twist angle between the two materials are determined from the constant binding energy cuts shown in Figs. \ref{fig:1}(f)-(h). From the relative orientation of the hole pockets, we estimate a twist angle of (23 $\pm$ 1)$^{\circ}$. The energy distribution curves (EDCs) in Fig. \ref{fig:1}(i)-(j) track the VBM binding energy positions of the two materials.
The upper VB of SL WS$_2$ is located inside the band gap of h-BN and the SL WS$_2$ VBM is characterized by a spin-orbit splitting of 430~meV (see Fig. \ref{fig:1}(j)), in agreement with theoretical predictions \cite{Kormnyos:2015} and previous experiments \cite{Dendzik:2015,Ulstrup:2016}. The clear electronic states and lack of band hybridization reveal a weak interlayer interaction between the two materials.
Similar to work on graphene/h-BN \cite{Wang:2016b}, we expect these data to represent the intrinsic electronic structure of SL WS$_2$ with negligible substrate influence.

The impact of electron doping on the electronic structure via \textit{in-situ} surface potassium deposition is shown in Fig.~\ref{fig:2}(a,b) (see Supplementary Figure S2 for core level data on clean and potassium dosed samples). Doping WS$_2$/h-BN leads to the CBM being populated at the \kbar\ points of SL WS$_2$, confirming the expected direct band gap. A surprising change of the dispersion of the two spin-orbit split bands VB$_A$ and VB$_B$ of WS$_2$/h-BN around the \kbar\ point is observed, highlighted in the EDCs in Fig.~\ref{fig:2}(c). The spin-splitting due to spin-orbit coupling $\Delta_{SO}$ increases from 430~meV in the undoped case to 660~meV in the electron-doped case, as sketched in Fig.~\ref{fig:2}(d). Such a large spin-splitting has not previously been observed in any SL material to our knowledge. In this case, the band gap of SL WS$_2$ is 1.65~eV and the CBM to VB$_B$ offset is 2.31~eV. These values are sketched in Fig.~\ref{fig:2}(d) and denoted as $E_A$ and $E_B$, respectively, due to the relation with the A and B exciton lines observed in optical experiments \cite{Qiu:2013}. 
\figthree

A detailed evolution of the band extrema with increasing doping is shown in Fig.~\ref{fig:3}(a-e).
The dispersions around the VBM as determined from EDC line shape analysis (see Supplementary Figure S6) are shown by dashed red curves and directly compared in panel (f). We estimate the charge carrier density, $N$, at each dosing step from the CBM position (see methods). These estimated doping levels  are consistent with those achieved in similar experiments on bulk WSe$_2$ \cite{Riley:2015} and with the intensity of the potassium 3p core level (see Supplementary Figure S3).

From EDC peak positions at \kbar\ (see Supplementary Figure S7), we extract the VBM and CBM energies as a function of dosing (panel~(g)). After the first dosing step ($N=1.7\times 10^{13}$~cm$^{-2}$), the CBM becomes occupied and VB$_A$ and VB$_B$ rigidly shift to higher binding energy. With further dosing, the CBM moves further down to higher binding energy, while VB$_A$ and VB$_B$ shift back towards lower binding energy, resulting in a continuous narrowing of the band gap. In particular, the dispersion of VB$_A$ appears to renormalize with increased doping (see panels (d,e,f)), with a dramatic increase in $\Delta_{SO}$ (panels~(g,h)). This leads to a corresponding change in the relative energy separation between $E_A$ and $E_B$ (panel~(i)), implying that the energies of the A and B exciton lines also separate. The data points with different marker shape and color in panels~(h,i) stem from separate doping experiments on the three different flakes studied in Fig. 2, Fig. 3 and in Supplementary Figure S8. A reproducible trend is found across all flakes. Note also that in the carrier density range between $2\times10^{12}$~cm$^{-2}$ and $1.0\times 10^{13}$~cm$^{-2}$ we find a more modest band gap renormalization of (90 $\pm$ 30)~meV, which is in excellent agreement with gated device measurements on a similar sample \cite{Chernikov:2015b}. Our observations reveal that it is insufficient to only consider rigid band shifts, and that strong dispersion changes can result from doping of SL TMDs.

The surprising doping-induced changes in $\Delta_{SO}$ are likely not directly related to the surface potential induced by the potassium deposition (through, \textit{e.g.}, the Rashba interaction) which is not expected to affect $\Delta_{SO}$ at \kbar~for SL TMDs. This rather introduces a splitting at \gbar, which we do not observe \cite{Yuan:2013,Shanavas:2015}. Furthermore, we can rule out any potassium induced structural symmetry breaking in our heterostructure, as only minor rigid binding energy shifts of the S 2p core levels of WS$_2$ and of the underlying h-BN $\pi$-band are observed after complete doping (see Supplementary Figures S4-S5). The reproducible charge carrier dependence of the spectral changes demonstrated in Figs. 2-3 and Supplementary Figure S8 suggest that these changes originate from Coulomb interactions around the band extrema of SL WS$_2$ \cite{Chernikov:2015b,Liang:2015,Ferreiros:2014}.

The linewidths of the VB$_A$ and CBM peaks exhibit a non-monotonic dependence with doping, which can not be described by simple scattering on ionized potassium impurities (see further discussion in Supplementary Section 4). Specifically, the observation that VB$_A$ renormalization coincides with occupation of the CBM suggests that the renormalization is caused by new scattering channels available upon occupation of the conduction band (CB). Previous works utilizing surface potassium deposition for electron-doping of SL TMDs on conductive substrates \cite{Zhang:2014a,Zhang:2016,Miwa:2015} have shown no such changes in $\Delta_{SO}$, where the Coulomb interactions are already strongly screened in the undoped case \cite{Ugeda:2014}. We believe that the reduced dielectric constant of the h-BN substrate plays a key role for these observations as it leads to reduced screening of the many-body interactions in the bare SL WS$_2$.
\figfour

An alkali-atom-induced renormalization of the VB edge at K towards \EF, observed in several bulk TMDs, has been attributed to the breaking of the outermost layers' degeneracy by the doping-induced field \cite{Riley:2015,Kang:2017}. This can either be a single-particle effect \cite{Kang:2017} or a combination of single- and many-body effects \cite{Riley:2015}, the latter of which suggests a negative electronic compressibility (NEC), the motion of the chemical potential $\mu$ towards the VBM, i.e. $d\mu/dN<0$.

In contrast, we observe distinctive effects in SL WS$_2$/h-BN, namely 1) a renormalization of $\Delta_{SO}$ within the single layer, 2) an NEC in which $|d\mu/dN|$ is significantly larger than in the bulk \cite{Riley:2015,Kang:2017}, and 3) the VB$_A$ slope is  discontinuous at  $k=($\kbar,~\kbarp$) ~\pm \sim$0.15~\AA$^{-1}$ in Fig.\ \ref{fig:2}(b) at high doping. This leads to kinks in VB$_A$, exemplified by the arrow in Fig.\ \ref{fig:3}(e), which develop continuously in strength with doping in Figs.\ \ref{fig:3}(a)-(e). Such kinks are common in ARPES when the created ``photohole'' interacts strongly with well-defined (in energy and/or momentum) excitations \cite{Bostwick:2007}.

As noted above, the band renormalization coincides with the occupation of the CB, suggesting that such excitations are associated with electron-hole ($e$-$h$) pairs near \EF~in the CB, induced in response to the VB hole created during photoemission. In the undoped situation sketched in Fig. \ref{fig:4}(a) such interactions are not possible. At high carrier densities where the CB is occupied an $e$-$h$ excitation around \kbar\ (or \kbarp) can interact with the VB photohole, forming a positively charged,  bound electron-hole-hole complex denoted as $e_ch_ch_v$, where $(c,v)$ denotes charges in the (CB,VB), respectively, as illustrated in Fig. \ref{fig:4}(b). Such a process would renormalize the bare band dispersion and lifetime of the VB states, broadening and shifting their spectra as observed.

These excitations may be compared to the \xplus\ (\xminus) trions found in $p$- ($n$)-doped TMDs with configuration $e_ch_vh_v$ ($e_ce_ch_v$). Such trions have been invoked to interpret additional spectral lines shifted in energy by $\sim$~20-60~meV with respect to the main A exciton line in optical absorption \cite{Mak:2013} and luminescence \cite{Mak:2013,Zhu:2015} measurements of SL TMDs. Our measurements show a relative shift of VB$_A$ by $\sim$ 0.23 eV compared to VB$_B$, which reflects the absolute binding energy reduction $\Delta E_{ehh}$ of the photohole associated with the formation of the trion (see Fig.~\ref{fig:4}(b)). The order of magnitude of $\Delta E_{ehh}$ is compatible with the absolute trion binding energies that can be extracted from the optical experiments \cite{Mak:2013,Zhu:2015}, however, the exact values are expected to depend on the dielectric environment of the sample and the doping. So far, corresponding trion features associated with B excitons have not been observed in optical experiments, which is fully consistent with the absence of renormalization of the VB$_B$ dispersion in the present ARPES data, as seen in Figs. \ref{fig:3}(a)-(e) and as sketched in Fig. \ref{fig:4}(c). This lack of renormalization of the VB$_B$ dispersion might be attributed to additional decay channels of the VB$_B$ hole such as decay into the VB$_A$ band. The dramatic increase of $\Delta_{SO}$ and the band gap renormalization can therefore be viewed as direct consequences of forming trionic quasiparticles around the VB$_A$ and CB extrema.

The assignment of trions in optical measurements of semiconducting TMDs is currently being debated, as a recent theoretical study points towards the possibility of interactions between $e$-$h$ pairs and the remaining charge density forming other types of quasiparticles such as exciton-polarons \cite{Efimkin:2017}.  Our ARPES measurements provide direct evidence for such multi-component excitations in SL TMDs and gives access to both their energy and momentum dependence that is lacking from momentum-integrating transport, optical, or tunneling measurements. We envision further theoretical and experimental studies to disentangle such many-body effects in the spectral function of SL TMDs. The charge carrier dependent electronic band gap and spin-splitting that arise from these many-body effects will profoundly impact the charge-, spin- and valley-dependent dynamics and transport properties of devices, as well as the interpretation of excitonic effects.  

\section{Methods}
\textbf{Fabrication of WS$_2$/h-BN heterostructures.} The heterostructures were prepared by successively transferring few layer h-BN (commercial crystal from HQ Graphene) and then SL WS$_2$ onto 0.5 wt \% Nb-doped rutile TiO$_2$(100) purchased from Shinkosha Co., Ltd. A thin film of polycarbonate (PC) was mounted onto polydimethylsiloxane (PDMS) on a glass slide to prepare a PC/PDMS stamp. This stamp was first utilized to pick up h-BN flakes from a SiO$_2$ substrate and then dropped onto the TiO$_2$ substrate under a microscope. The transferred h-BN flakes were cleaned of any polymer residue by annealing at 625~K in UHV for 1 hour. Next, SL WS$_2$ flakes were picked up from the SiO$_2$ growth substrate and aligned to drop onto h-BN, such that a part of the flake makes contact with the TiO$_2$ substrate. The process is followed by another annealing step in UHV to clean of any remaining residues.

\textbf{microARPES experimental details.} The samples were transported through air to the Microscopic and Electronic Structure Observatory (MAESTRO) at the Advanced Light Source (ALS) where they were inserted in the microARPES UHV end-station with a base pressure better than $5 \times 10^{-11}$~mbar. The samples were given a mild anneal at 600~K prior to measurements in order to desorb adsorbates from air. The synchrotron beam-spot size was on the order of 10~$\mu$m for the photon energies of 145~eV and 76.5~eV used to obtain the microARPES data. The data were collected using a hemispherical Scienta R4000 electron analyzer equipped with custom-made deflectors that enable collecting ARPES spectra over a full BZ without moving the sample. Potassium dosing experiments were carried out \textit{in situ} using SAES getters mounted in the analysis chamber such that dosing could be completed on an optimum sample position without ever moving the sample. Core level data of undoped and potassium dosed samples are presented in Supplementary Figure S2 to document the cleanliness of the samples. The data in Figs. \ref{fig:1}-\ref{fig:2} are from the same sample, while the data in Fig. \ref{fig:3} and Supplementary Figure S8 were obtained on fresh samples. The charge carrier density $N$ is estimated using a simple model of a 2D parabolic band given by $N = (g_vg_sm^{\ast}_{cb}k_BT/2\pi\hbar^2)\ln(1+e^{E-E_F/k_BT})$, where the factors $g_v = 2$ and $g_s = 2$ take spin- and valley-degeneracy into account, $m^{\ast}_{cb}$ is the effective mass of the SL WS$_2$ CB obtained from Ref. \cite{Kormnyos:2015}, $k_B$ is Boltzmann's constant, $\hbar$ is the reduced Planck constant, $T$ is the sample temperature and $E-E_F$ is determined from the fitted CBM position. The total energy and momentum resolution in the microARPES data were better than 20~meV and 0.01~\AA$^{-1}$, respectively. Measurements and dosing experiments were carried out at both 85~K and at 20~K, without any noticeable change in behavior between the two temperatures.

\textbf{Data availability.} The data that support the plots within this paper and other findings of this study are available from the corresponding author upon reasonable request.

\section{acknowledgement}
We thank Allan H. MacDonald for helpful discussions. S. U. acknowledges financial support from the Danish Council for Independent Research, Natural Sciences under the Sapere Aude program (Grant No. DFF-4090-00125) and from VILLUM FONDEN (Grant. No. 15375). R. J. K. is supported by a fellowship within the Postdoc-Program of the German Academic Exchange Service (DAAD). S. M. acknowledges support by the Swiss National Science Foundation (Grant No. P2ELP2-155357). The Advanced Light Source is supported by the Director, Office of Science, Office of Basic Energy Sciences, of the U.S. Department of Energy under Contract No. DE-AC02-05CH11231. This work was supported by IBS-R009-D1. The work at Ohio State was primarily supported by NSF-MRSEC (Grant DMR-1420451). Work at NRL was supported by core programs and the NRL Nanoscience Institute, and by AFOSR under contract number AOARD 14IOA018- 134141. 

\section{Author Contributions}
J. K. and S. U. conceived and planned the experiments. 
K. M. M. and B. T. J. synthesized the SL WS$_2$ flakes on SiO$_2$. 
J. K., S. S., J. X. and R. K. K assembled the WS$_2$/h-BN heterostructures on TiO$_2$.
S. U., R. J. K., S. M., J. K., A. B., E. R. and C. J. performed the microARPES experiments.
The microARPES setup was developed and maintained by C. J., A. B. and E. R..
S. U. analyzed the experimental data with inputs from C. J. and E. R..
All authors contributed to the interpretation and writing of the manuscript.

\section{Author information}
The authors declare that they have no competing financial interests.
Supplementary Information accompanies this paper.
Correspondence and requests for materials should be addressed to C.J. (CMJozwiak@lbl.gov)

\end{document}